\makeindex \textwidth 160mm \textheight 230mm \topmargin -5mm
\newcommand{\be}{\begin{equation}}
\newcommand{\ee}{\end{equation}}
\newcommand{\bea}{\begin{eqnarray}}
\newcommand{\eea}{\end{eqnarray}}

\def\asec{$''$ cy$^{-1}$}

\def\bb{\bibitem}
\def\rfr#1{eq.(\ref{#1})}

\def\eqi{\begin{equation}}
\def\eqf{\end{equation}}
\documentclass[11pt]{article}
\usepackage{amsmath,amsthm,amscd,amssymb}
\usepackage{latexsym}
\usepackage{graphicx,epsfig}

\begin{document}

\noindent{\bf \LARGE{On the perspectives of testing the
Dvali-Gabadadze-Porrati gravity model with the outer planets of
the Solar System}}
\\
\\
\\
{L. Iorio, }\\
{\it Viale Unit$\grave{a}$ di Italia 68, 70125\\Bari, Italy
\\e-mail: lorenzo.iorio@libero.it}\\\\
{G. Giudice, }\\
{\it  Dipartimento di Progettazione e Gestione Industriale,
Piazzale Tecchio 80, 80125\\Napoli, Italy}

\begin{abstract}
The multidimensional braneworld gravity model by Dvali, Gabadadze
and Porrati was primarily put forth to explain the observed
acceleration of the expansion of the Universe without resorting to
dark energy. One of the most intriguing features of such a model
is that it also predicts small effects on the orbital motion of
test particles which could be tested in such a way that local
measurements at Solar System scales would allow to get information
on the global properties of the Universe. Lue and Starkman derived
a secular extra-perihelion $\omega$ precession of $5\times
10^{-4}$ arcseconds per century, while Iorio showed that the mean
longitude $\lambda$ is affected by a secular precession of about $
10^{-3}$ arcseconds per century. Such effects depend only on the
eccentricities $e$ of the orbits via second-order terms: they are,
instead, independent of their semimajor axes $a$. Up to now, the
observational efforts focused on the dynamics of the inner planets
of the Solar System whose orbits  are the best known via radar
ranging. Since the competing Newtonian and Einsteinian effects
like the precessions due to the solar quadrupole mass moment
$J_2$, the gravitoelectric and gravitomagnetic part of the
equations of motion reduce with increasing distances, it would be
possible to argue that an analysis of the orbital dynamics of the
outer planets of the Solar System, with particular emphasis on
Saturn because of the ongoing Cassini mission with its precision
ranging instrumentation,  could be helpful in evidencing the
predicted new features of motion. In this paper we investigate
this possibility by comparing both analytical and numerical
calculation with the latest results in the planetary ephemeris
field. Unfortunately, the current level of accuracy rules out this
appealing possibility.
\end{abstract}


\section{Introduction}
Recently, Dvali, Gabadadze and Porrati (DGP) put forth a model of
gravity \cite{DGP00} which allows to explain the observed
accelerated expansion of our Universe without resorting to the
concept of dark energy. In such a picture our Universe is a (3+1)
space-time brane embeddeed in a 5-dimensional Minkowskian bulk.
While the dynamics of the Standard Model particles and fields is
confined to our brane, gravity can fully explore the entire bulk
getting strongly modified at large distances of the order of
$r_0\sim 5$ Gpc. An intermediate regime is set by the Vainshtein
scale  $r_{\star}=(r_g r_0^2)^{1/3}$, where $r_g=2GM/c^2$ is the
Schwarzschild radius of a central object acting as source of
gravitational field. For a Sun-like star $r_{\star}$ amounts to
about 100 parsec. An updated overview of the phenomenology of DGP
gravity can be found in \cite{Lue05}.
\subsection{Local orbital effects}
One of the most interesting features of such a picture is that in
the process of recovering the 4-dimensional Newton-Einstein
gravity for $r<<r_{\star}<< r_0$, DGP predicts small deviations
from it which yield to effects observable at local scales
\cite{DGZ03}. They come from an extra radial acceleration  of the
form \cite{Gruz05, LS03, Ior05b} \eqi \boldsymbol a_{\rm
DGP}=\mp\left(\frac{c}{2r_0}\right)\sqrt{ \frac{GM}{r}
}\hat{\boldsymbol r}.\label{acc} \eqf The minus sign is related to
a cosmological phase in which, in absence of cosmological constant
on the brane, the Universe decelerates at late times, the Hubble
parameter $H$ tending to zero as the matter dissolves on the
brane: it is called Friedmann-Lema$\hat{\rm
\i}$tre-Robertson-Walker (FLRW) branch. The plus sign is related
to a cosmological phase in which the Universe undergoes a deSitter
expansion with the Hubble parameter $H=c/r_0$ even in absence of
matter. This is the self-accelerated branch, where the accelerated
expansion of the Universe is realized without introducing a
cosmological constant on the brane. Thus, there is a very
important connection between local and cosmological features of
gravity in the DGP model. About the local effects, Lue and
Starkman \cite{LS03} derived an extra-secular precession of the
pericentre $\omega$ of a nearly circular orbit of a test particle
of $5\times 10^{-4} $ arcseconds per century (\asec), while Iorio
\cite{Ior05b} showed that also the mean anomaly $\mathcal{M}$ is
affected by DGP gravity at a larger extent; the longitude of the
ascending node $\Omega$ is left unchanged. As a result, the mean
longitude $\lambda=\omega+\Omega+\mathcal{M}$, which is a widely
used orbital parameter for nearly equatorial and circular orbits
as those of the Solar System planets, undergoes a secular
precession of the order of $10^{-3}$ \asec. It is independent of
the semi-major axis $a$ of the planetary orbits and depends only
on their eccentricities $e$ via second-order terms. Recent
improvements in the accuracy of the data reduction process for the
inner planets of the Solar System \cite{Pit05a, Pit05b}, which can
be tracked via radar-ranging,  have made the possibility of
testing DGP very thrilling \cite{Ior05a, Ior05b, Ior05c}. In
particular, Iorio \cite{Ior05c} showed that the recently observed
secular increase of the Astronomical Unit \cite{KraBru04, Sta05}
can be explained by the self-accelerated branch of DGP and that
the predicted values of the Lue-Starkman perihelion precessions
for the self-accelerated branch are compatible with the recently
determined extra-perihelion advances \cite{Pit05b}, especially for
Mars, although the errors are still large.
\subsection{Problems of DGP gravity}
There are still some issues about the theoretical consistency of
the DGP model, so that direct, local observational tests would be
very useful and important to tackle observationally this matter.
The recovery of the Newton-Einstein four dimensional gravity is
rather not trivial. The aforementioned orbital effects come from
the correction to the Newtonian potential of a Schwarzschild
source found in \cite{Gruz05, LS03}. Such a potential is obtained
within a certain approximation which is valid below the Vainshtein
scale. However, it is not yet clear, at present, whether this
potential can match continuously onto a four-dimensional Newtonian
potential above the Vainshtein scale, and then, also match onto
the five-dimensional potential above the crossover scale $r_0$. An
alternative solution that smoothly interpolates between the
different regions was discussed in \cite{GaIg05a,GaIg05b}. The
correction to the Newtonian potential arising from that solution
below the Vainshtein scale is somewhat different from what used
here. In particular, it is reduced by a multiplicative factor
smaller than unity. As a consequence, the predictions are also
different. Moreover, according to the authors of \cite{ghosts},
ghosts affect the self-accelerated branch, plaguing the
consistency of the model at cosmological scales. In regard to the
domain below the Vainshtein scale, according to \cite{ghosts2} a
ghost appears in the Pauli-Fierz model of massive gravity, while
it is absent in DGP on the conventional branch. A similar result
has also been obtained in \cite{gh3}.

%
\subsection{Aim of the paper}
As already pointed out, up to now the only available preliminary
tests of the DGP gravity refer to the inner planets of the Solar
System. In this paper we focus on the outer planets  in order to
enlarge the possible set of independent checks. For such celestial
bodies many competing Newtonian and Einsteinian effects are
smaller than the DGP features of motion allowing, in principle,
cleaner tests of the Lue-Starkman and Iorio precessions.

\section{Can the outer planets of the Solar System be useful?}
\subsection{The DGP precessions and the competing Newtonian and Einsteinian effects}
%
 We
list the DGP precessions and the competing Newtonian and
Einsteinian precessions in Table \ref{sensi}, taken from
\cite{Ior05b}, for the mean longitudes.
\begin{table}
\caption{Nominal values, in \asec, of the secular precessions
induced on the planetary mean longitudes $\lambda$ by the DGP
gravity and by some of the competing Newtonian and Einsteinian
gravitational perturbations. For a given planet, the precession
labelled with Numerical includes all the numerically integrated
perturbing effects of the dynamical force models used at JPL for
the DE200 ephemeris. E.g., it also comprises the classical N- body
interactions, including  the Keplerian mean motion $n$. For the
numerically integrated planetary precessions see
http://ssd.jpl.nasa.gov/elem$\_$planets.html$\#$rates. The effect
labelled with GE is due to the post-Newtonian general relativistic
gravitoelectric Schwarzschild component of the solar gravitational
field \cite{Einstein 1915}, that labelled with $J_2$ is due to the
Newtonian effect of the Sun's quadrupole mass moment $J_2$
\cite{J2} and that labelled with LT is due to the post-Newtonian
general relativistic gravitomagnetic Lense-Thirring \cite{Lense
and Thirring 1918} component of the solar gravitational field (not
included in the force models adopted by JPL). For $J_2$ the value
$1.9\times 10^{-7}$ has been adopted \cite{Pit05b}. For the Sun's
proper angular momentum $S$, which is the source of the
gravitomagnetic field, the value $1.9\times 10^{41}$ kg m$^2$
s$^{-1}$ \cite{Pijpers 2003} has been adopted. }\label{sensi}
\begin{tabular}{@{\hspace{0pt}}llllll}
\hline\noalign{\smallskip} Planet & DGP & Numerical & GE & $J_2$ &
LT
\\
\noalign{\smallskip}\hline\noalign{\smallskip} Mercury &
$1.2\times 10^{-3}$ & $5.381016282\times 10^8$ & $-8.48\times
10^1$ & $4.7\times 10^{-2}$ & $-2\times
10^{-3}$\\
Venus & $1.2\times 10^{-3}$ & $2.106641360\times 10^8$&
$-1.72\times
10^1$ &$5\times 10^{-3}$ & $-3\times 10^{-4}$\\
Earth & $1.2\times 10^{-3}$ & $1.295977406\times 10^8$ & $-7.6$ &
$1.6\times 10^{-3}$& $-1\times 10^{-4}$\\
Mars & $1.2\times 10^{-3}$ & $6.89051037\times 10^7$ & $-2.6$
&$3\times
10^{-4}$& $-3\times 10^{-5}$\\
Jupiter & $1.2\times 10^{-3}$ & $1.09250783\times 10^7$ &
$-1\times
10^{-1}$ & $5\times 10^{-6}$ & $-7\times 10^{-7}$\\
Saturn & $1.2\times 10^{-3}$ & $4.4010529\times 10^6$ & $-2\times
10^{-2}$ & $6\times 10^{-7}$ & $-1\times 10^{-7}$\\
Uranus & $1.2\times 10^{-3}$ & $1.5425477\times 10^{6}$ &$-4\times
10^{-3}$ & $5\times 10^{-8}$ & $-1\times 10^{-8}$\\
Neptune & $1.2\times 10^{-3}$ & $7.864492\times 10^{5}$ &
$-1\times
10^{-3}$ & $1\times 10^{-8}$ & $-5\times 10^{-9}$\\
Pluto & $1.2\times 10^{-3}$ & $5.227479\times 10^{5}$ & $-7\times
10^{-4}$ & $4\times 10^{-9}$ & $-2\times 10^{-9}$\\
 \noalign{\smallskip}\hline
\end{tabular}
\end{table}
In order to extract the DGP signal from the analysis of the
determined secular precessions of the orbital elements it is
mandatory that the competing effects are included in the dynamical
force models of the planetary data reduction softwares to  a
sufficient level of accuracy. It occurs for all such features of
motion apart from the Lense-Thirring effect, which is not modelled
at all in the currently used ephemeris softwares,  and the solar
quadrupole mass moment $J_2$ which is, in fact, modelled but whose
uncertainty is of the order of $10\%$. Such dynamical features
induce residual mismodelled precessions which could seriously
affect the recovery of a smaller effect. Otherwise, it could be
possible to suitably combine the orbital elements in order to a
priori cancel out some of the unwanted perturbations.

A careful inspection of Table \ref{sensi} shows that for the outer
planets of the Solar System-and for Mars-the nominal values  of
the LT and $J_2$ precessions are themselves well smaller than the
DGP rate. The nominal GE precessions are larger but by  a
sufficiently small amount to create no problems because they are
routinely included in the dynamical force models in terms of the
PPN parameters $\beta$ and $\gamma$ \cite{Will93} which are known
with an accuracy of $10^{-4}$ \cite{Pit05b} or better
\cite{Bert03}. Moreover, deviations from their relativistic values
are expected at the level of $10^{-6}-10^{-7}$ \cite{nordam}. The
Newtonian N-body precessions are known very well and the impact of
the asteroids, which limits the obtainable accuracy for the inner
planets with particular emphasis on Mars, is, of course,
negligible. So, at first sight, it would seem appealing to
consider such planets as ideal candidates to test DGP gravity
because of the absence of competing aliasing effects which could
mask it. This feeling is enforced by the ongoing Cassini tour of
the Saturn system which started at the end of 2004 and should last
for almost five years. Lue \cite{Lue05} hoped that the precise
radio-link to Cassini might, indeed, provide ideal test for the
anomalous periheion precession.
\subsection{Confrontation with data}
Unfortunately, the situation is in fact rather unfavorable for the
outer planets. Indeed, it must be pointed out that the
investigated effect is a secular one, i.e. it is integrated over
one orbital period of the planet. The currently available modern
observations for the outer planets span almost one century and
cover a limited number of full orbital revolutions of Jupiter
($P=11.86$ yr), Saturn ($P=29.46$ yr) and Uranus ($P=84$ yr). Up
to now, the orbit of Jupiter is the best known among the outer
planets because a number of precise radar observations by
spacecraft (Pioneer 10 and 11, Voyager 1 and 2, Ulysses and
Galileo) approaching the planet or orbiting it have been
performed. The other planets rely entirely upon optical
observations. The latest results by Pitjeva  \cite{Pit05a} are
reported in Table \ref{Pit}.
\begin{table}
\caption{Formal standard statistical errors in the non-singular
planetary orbital elements, from Table 4 of \cite{Pit05a}. The
latest EPM2004 ephemerides have been used. The angles $i$ and
$\varpi$ are the inclination and the longitude of perihelion,
respectively. The realistic errors may be up to one order of
magnitude larger. The units for the angular parameters are
milliarcseconds (mas).}\label{Pit}
\begin{tabular}{@{\hspace{0pt}}lllllll}
\hline\noalign{\smallskip} Planet & $a$ (m) & $\sin i\cos\Omega$
(mas)& $\sin i\sin\Omega$ (mas)& $e\cos\varpi$ (mas)&
$e\sin\varpi$ (mas) & $\lambda$ (mas)
\\
\noalign{\smallskip}\hline\noalign{\smallskip}
Mercury & 0.105 & 1.654 & 1.525 & 0.123 & 0.099 & 0.375\\
Venus & 0.329 & 0.567 & 0.567 & 0.041 &  0.043 & 0.187\\
Earth & 0.146 &- & -& 0.001 &  0.001 &  -\\
Mars & 0.657 &0.003 & 0.004 & 0.001 & 0.001 & 0.003\\
Jupiter & 639 & 2.410 & 2.207 & 1.280 &  1.170 & 1.109\\
Saturn & 4222 & 3.237 &  4.085 & 3.858 & 2.975 &  3.474 \\
Uranus & 38484 & 4.072 & 6.143 & 4.896 & 3.361 & 8.818\\
Neptune & 478532 & 4.214 & 8.600 & 14.066 & 18.687 & 35.163\\
Pluto & 3463309 & 6.899 & 14.940 &  82.888 & 36.700 & 79.089\\
 \noalign{\smallskip}\hline
\end{tabular}
\end{table}
It can be noted that, even in the case of Jupiter, the currently
available sensitivity does not allow to detect the DGP precession.
In regard to the mean longitudes of Jupiter and Saturn, their
(formal) uncertainties are slightly larger than the Iorio
precession. Another source of limiting systematic bias is
represented by the uncertainty $\delta n=(3/2)\sqrt{GM/a^5}\delta
a$ in the Keplerian mean motion $n=\sqrt{GM/a^3}$ induced by the
errors $\delta a$ in the semimajor axis. According to Table
\ref{Pit}, for Jupiter and Saturn we have $\delta n\sim 10^{-2}$
\asec\ which is one order of magnitude larger than the
investigated effect. The situation is worse for the perihelia:
indeed, the eccentricities of Jupiter and Saturn only amount to
0.04 and 0.05, respectively, so that their apsidal lines are very
difficult to be determined. As a consequence, from Table \ref{Pit}
it can be obtained an uncertainty up to 100 times larger than the
Lue-Starkman  perihelion precession. It is important to stress
that such evaluations are based on the formal, statistical errors
in the planetary orbital elements: realistic errors may be up to
ten times larger.
\subsubsection{Numerical simulations}
The Keplerian orbital elements are not directly measured
quantities: they are related to data in an indirect way. For the
outer planets the true observables are the right ascension
$\alpha$ and the declination $\delta$. Figure 2 of \cite{Pit05a}
shows the residuals, in $''$, of $\alpha\cos\delta$ and $\delta$
for Jupiter, Saturn, Uranus, Neptune and Pluto; the scale is $\pm
5$ $''$. In order to yield a direct and unambiguous confrontation
with the observations we decided to numerically produce a set of
residuals by integrating the planetary equations of motion with
and without the DGP acceleration of \rfr{acc} with the plus sign.


To integrate the equations of motion the Mercury package
\cite{Chamb99} has been chosen among the software packages freely
available on the Web because it is of simplest use and easily
allows to introduce user-defined perturbative forces, as the DGP
acceleration of \rfr{acc}. This package has been also used in
\cite{Iorio06}.

The Mercury package can use the following integration methods

\begin{itemize}
  \item second order mixed-variable symplectic
  \item Bulirsh-Stoer (general)
  \item Bulirsh-Stoer (conservative systems)
  \item Everhardt Radau 15th order
  \item hybrid (symplectic / Bulirsh-Stoer)
\end{itemize}

The perturbative force is introduced by giving the $x$, $y$, and
$z$ components in an heliocentric frame of the corresponding
acceleration.

Thanks to the experience gained during the integrations for
\cite{Iorio06}, the Everhardt Radau method has been chosen as
reference, due to its precision and speed. The initial epoch is
1913.0 (JD 2419770.5), and the initial planetary positions have
been obtained from JPL ephemerides.

\begin{figure}[htbp]
\begin{center}
\includegraphics[width=14cm,height=12cm,angle=0]{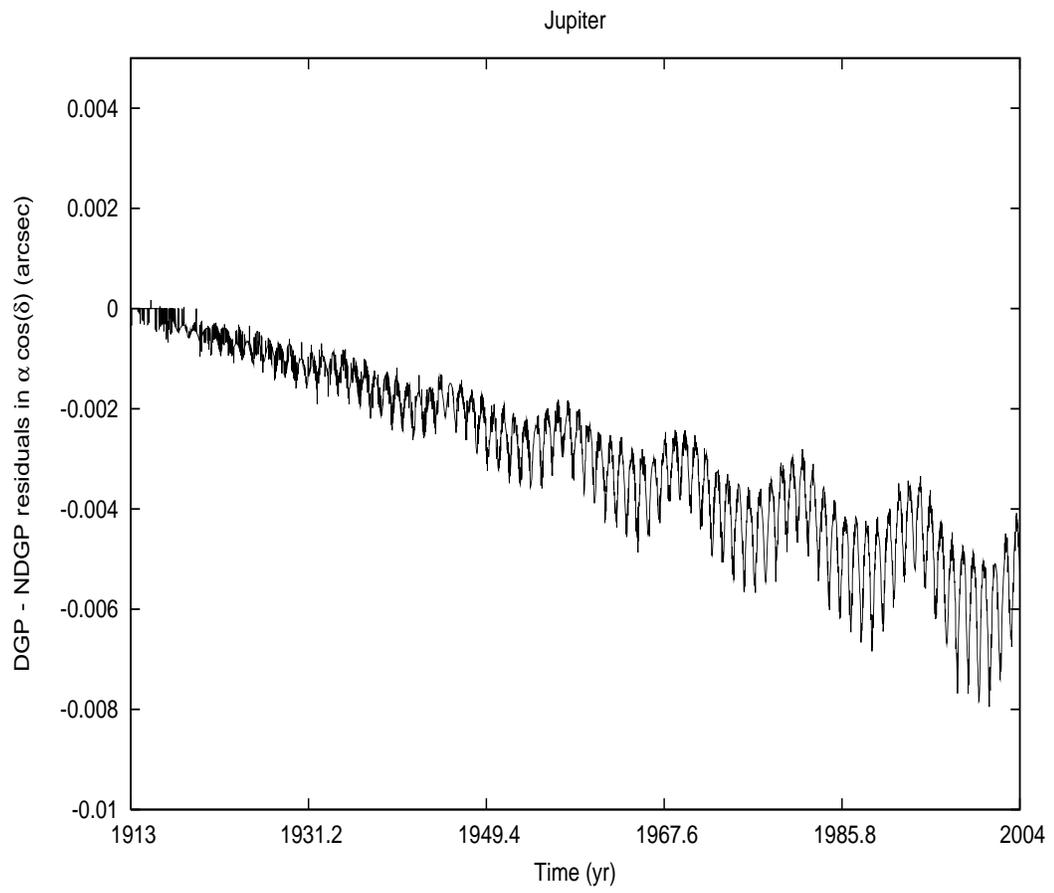}
\end{center}
\caption{\label{ju_ra} Jupiter: DGP shift, in arcseconds, for
$\alpha\cos\delta$ over $T=90$ years.}
\end{figure}

\begin{figure}[htbp]
\begin{center}
\includegraphics[width=14cm,height=12cm,angle=0]{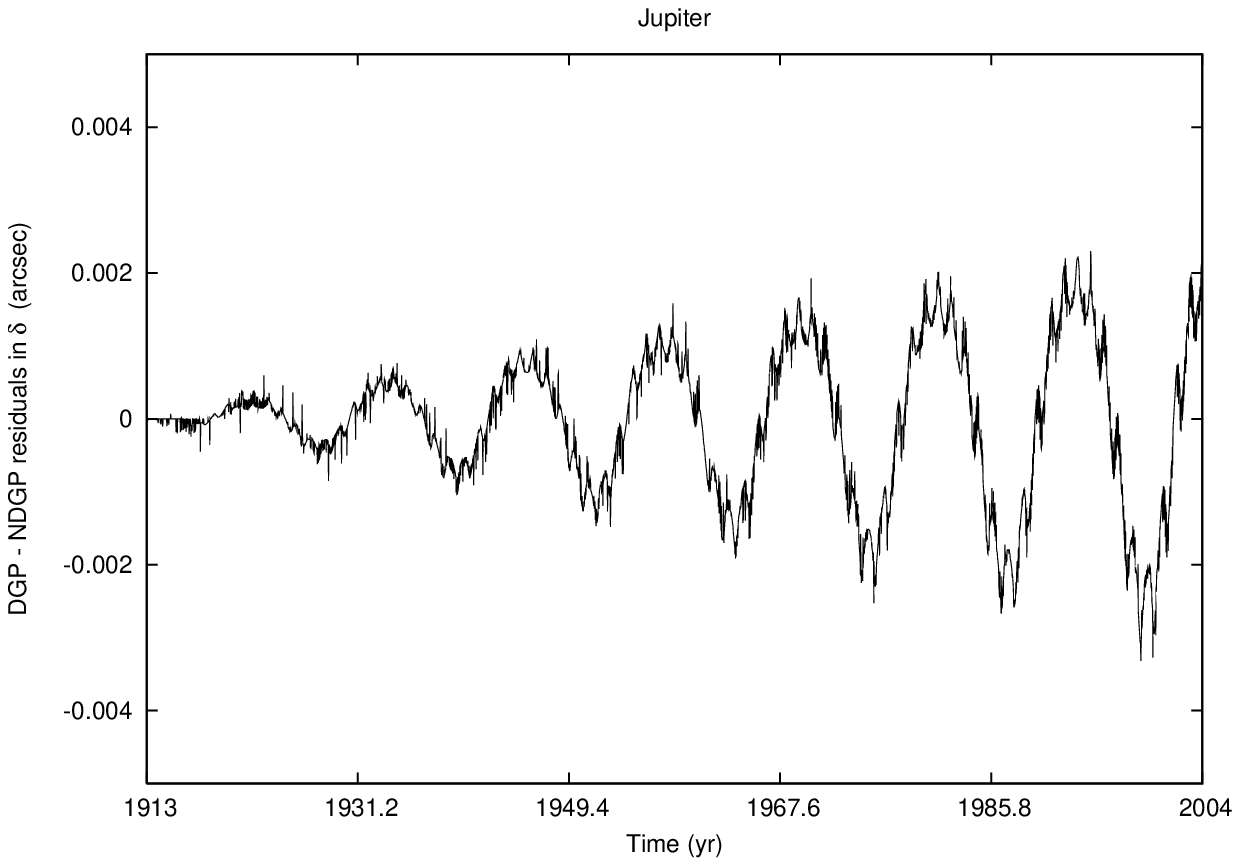}
\end{center}
\caption{\label{ju_de} Jupiter: DGP shift, in arcseconds, for
$\delta$ over $T=90$ years.}
\end{figure}

\begin{figure}[htbp]
\begin{center}
\includegraphics[width=14cm,height=12cm,angle=0]{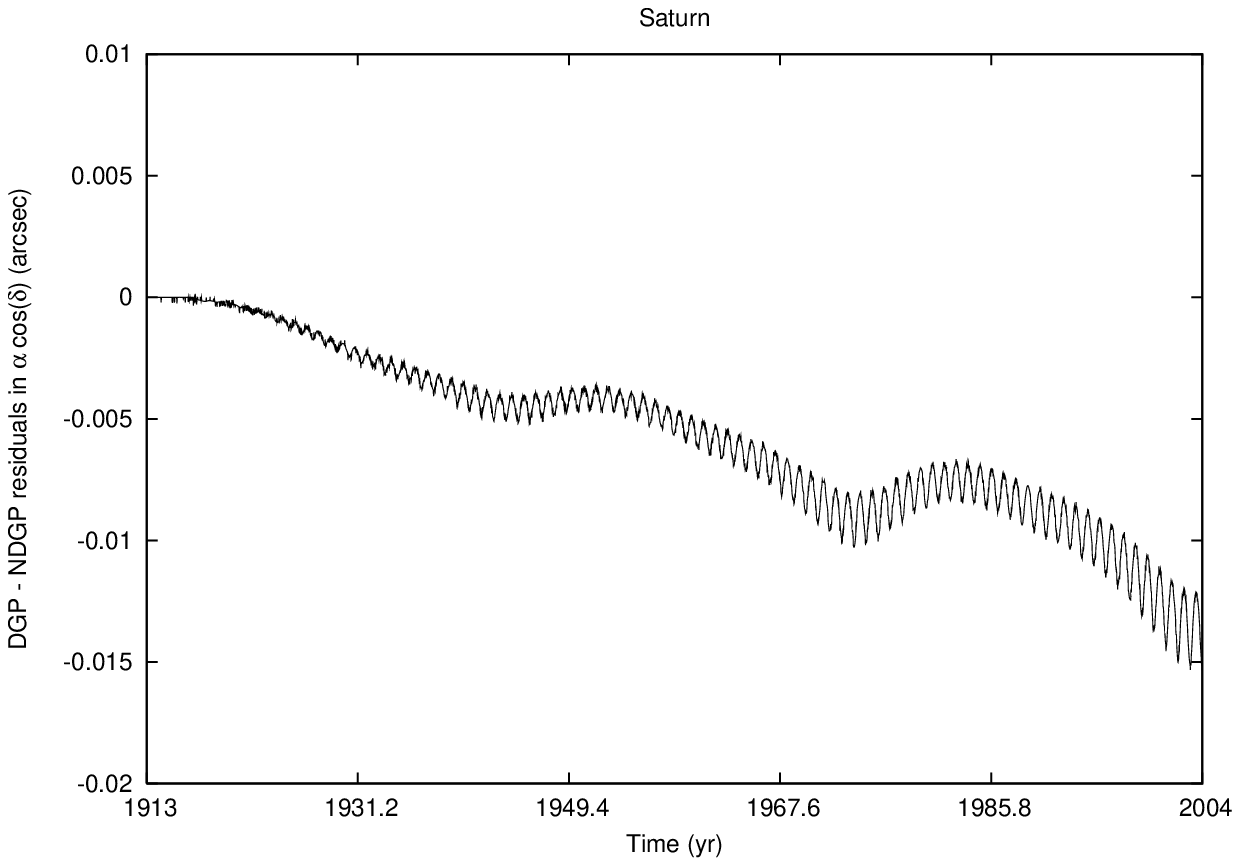}
\end{center}
\caption{\label{sa_ra} Saturn: DGP shift, in arcseconds, for
$\alpha\cos\delta$ over $T=90$ years.}
\end{figure}

\begin{figure}[htbp]
\begin{center}
\includegraphics[width=14cm,height=12cm,angle=0]{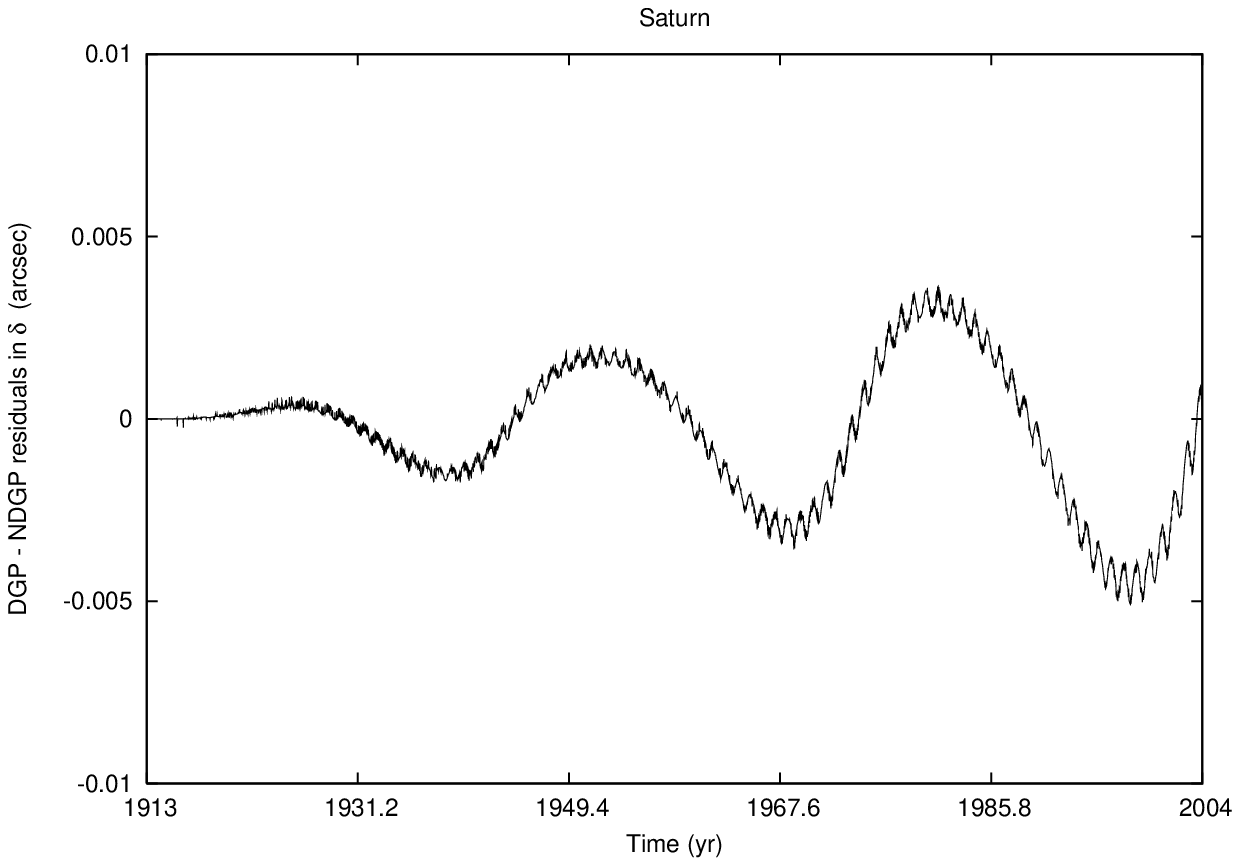}
\end{center}
\caption{\label{sa_de} Saturn: DGP shift, in arcseconds, for
$\delta$ over $T=90$ years.}
\end{figure}

\begin{figure}[htbp]
\begin{center}
\includegraphics[width=14cm,height=12cm,angle=0]{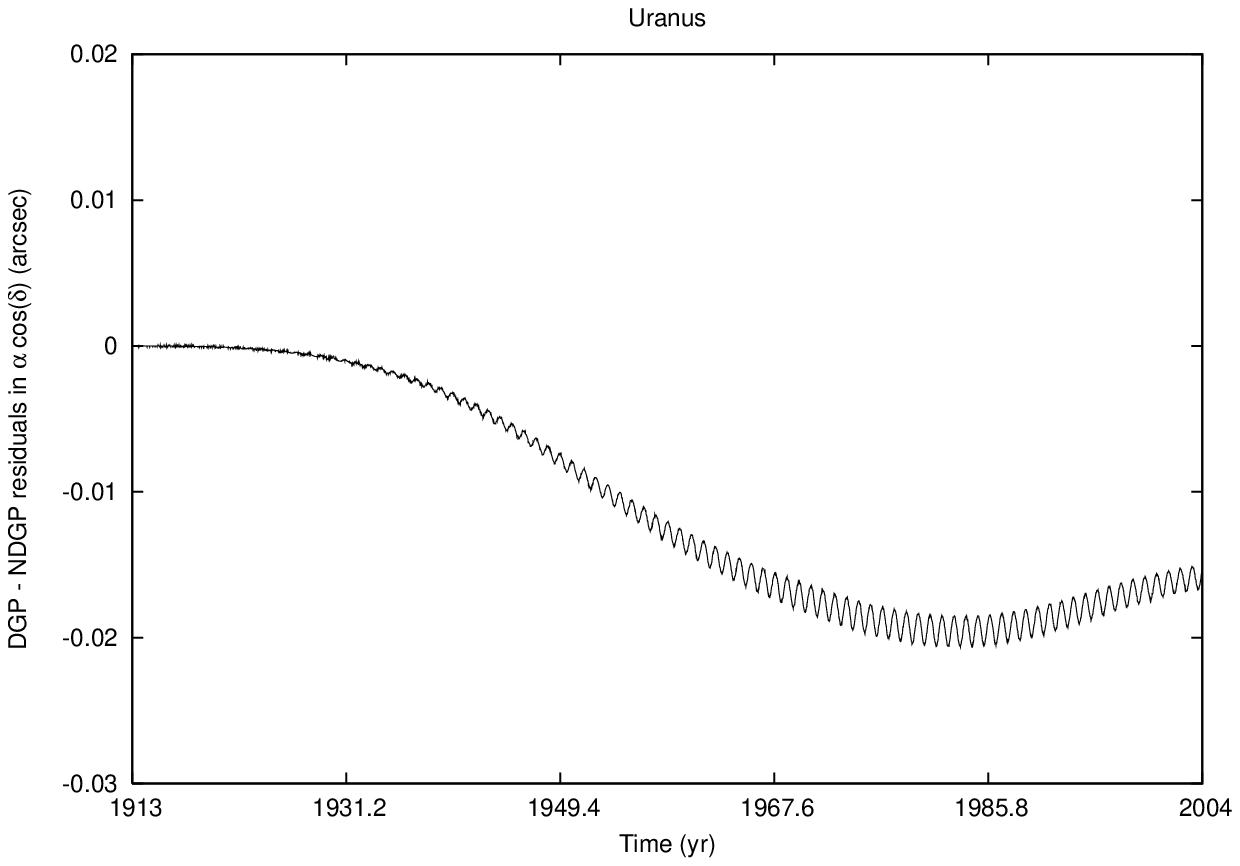}
\end{center}
\caption{\label{ur_ra} Uranus: DGP shift, in arcseconds, for
$\alpha\cos\delta$ over $T=90$ years.}
\end{figure}

\begin{figure}[htbp]
\begin{center}
\includegraphics[width=14cm,height=12cm,angle=0]{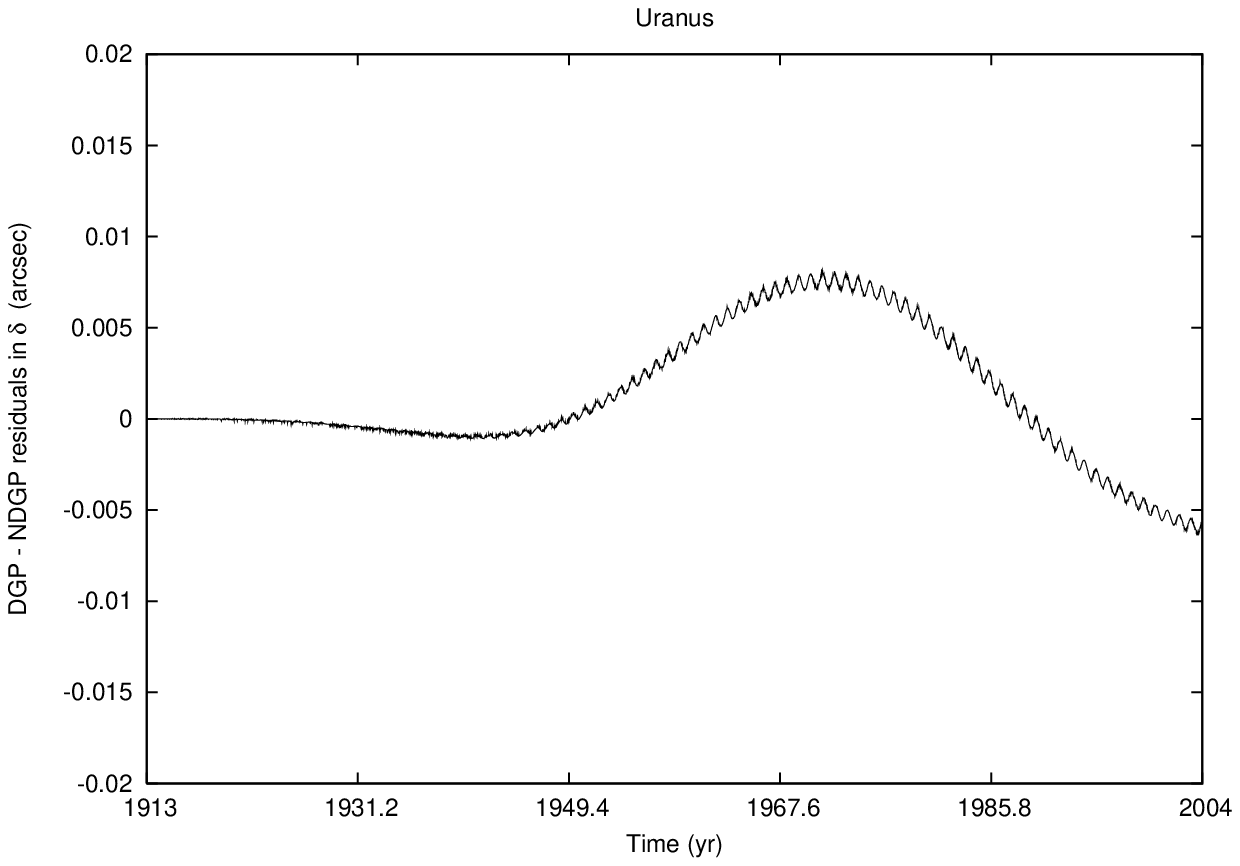}
\end{center}
\caption{\label{ur_de} Uranus: DGP shift, in arcseconds, for
$\delta$ over $T=90$ years.}
\end{figure}

\begin{figure}[htbp]
\begin{center}
\includegraphics[width=14cm,height=12cm,angle=0]{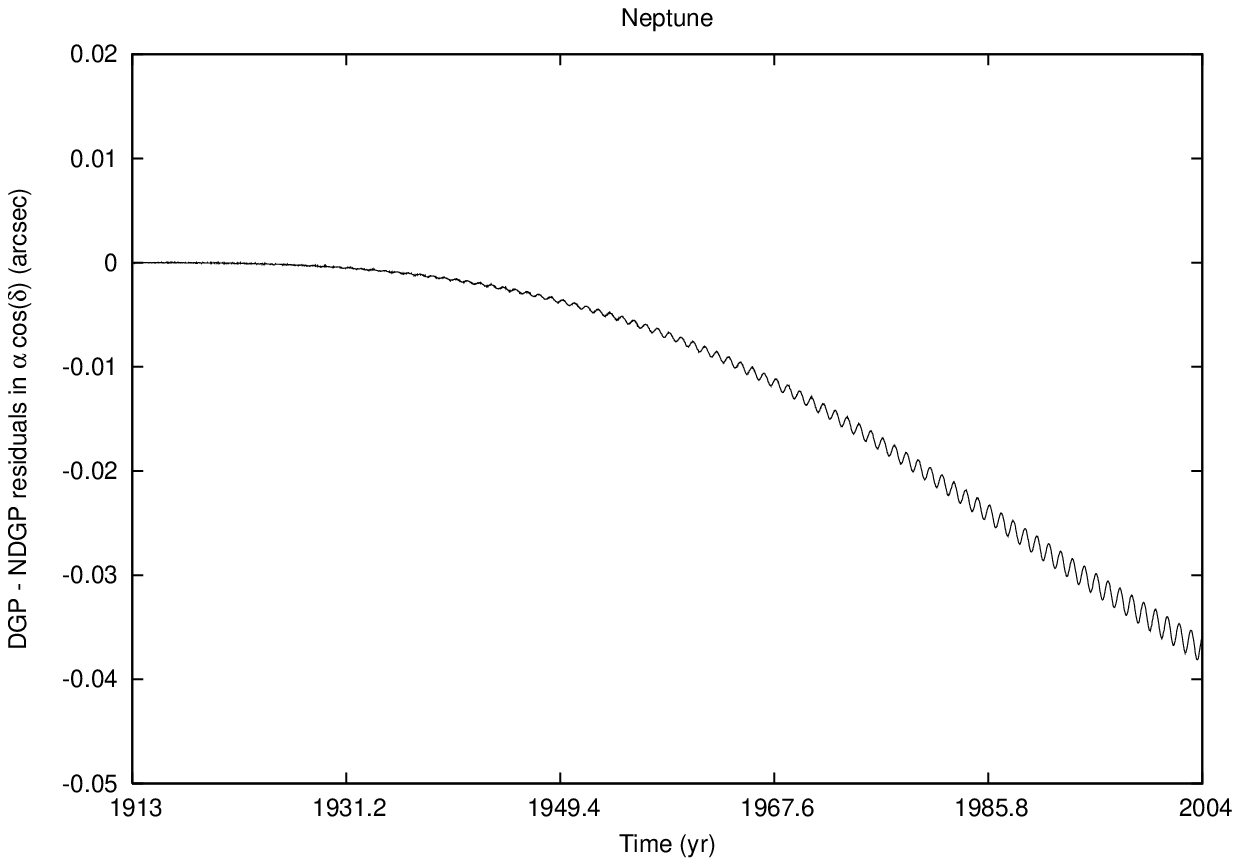}
\end{center}
\caption{\label{ne_ra} Neptune: DGP shift, in arcseconds, for
$\alpha\cos\delta$ over $T=90$ years.}
\end{figure}

\begin{figure}[htbp]
\begin{center}
\includegraphics[width=14cm,height=12cm,angle=0]{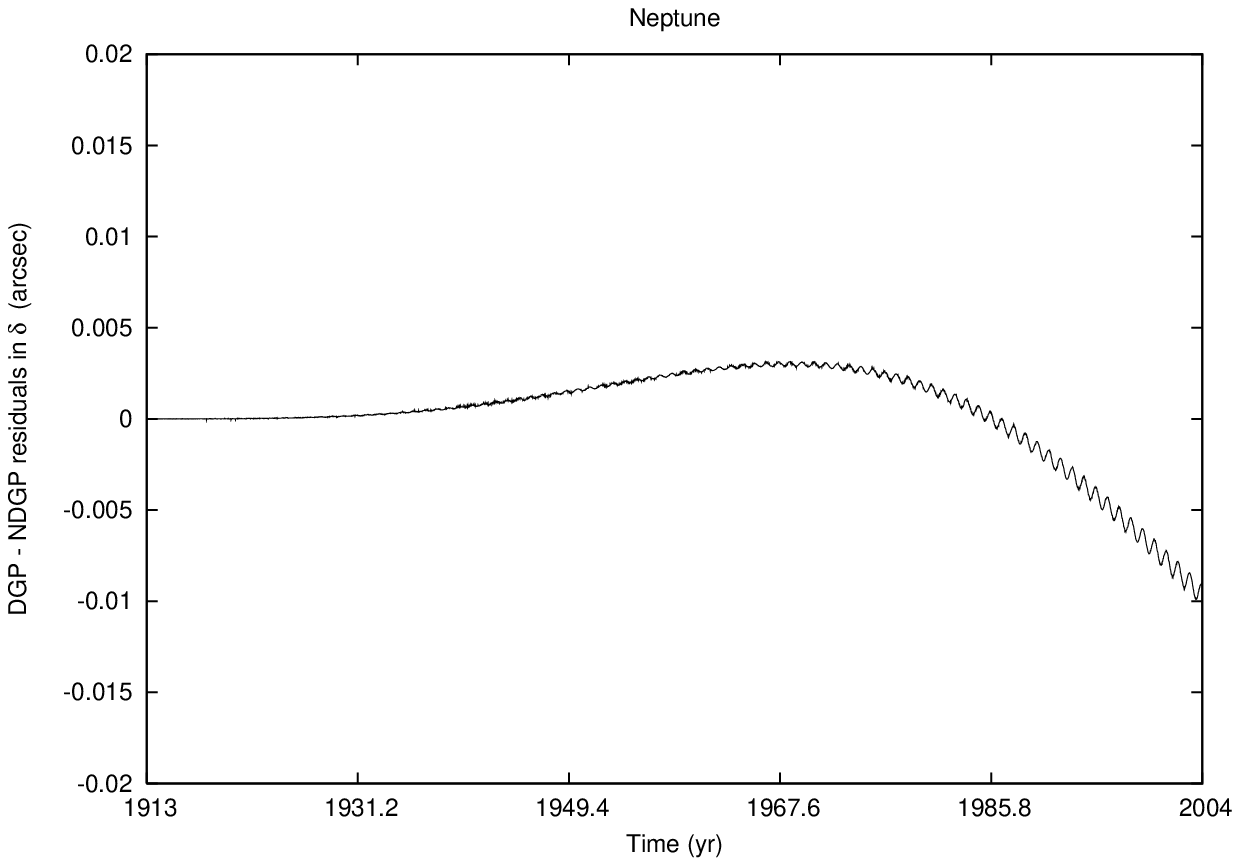}
\end{center}
\caption{\label{ne_de} Neptune: DGP shift, in arcseconds, for
$\delta$ over $T=90$ years.}
\end{figure}

\begin{figure}[htbp]
\begin{center}
\includegraphics[width=14cm,height=12cm,angle=0]{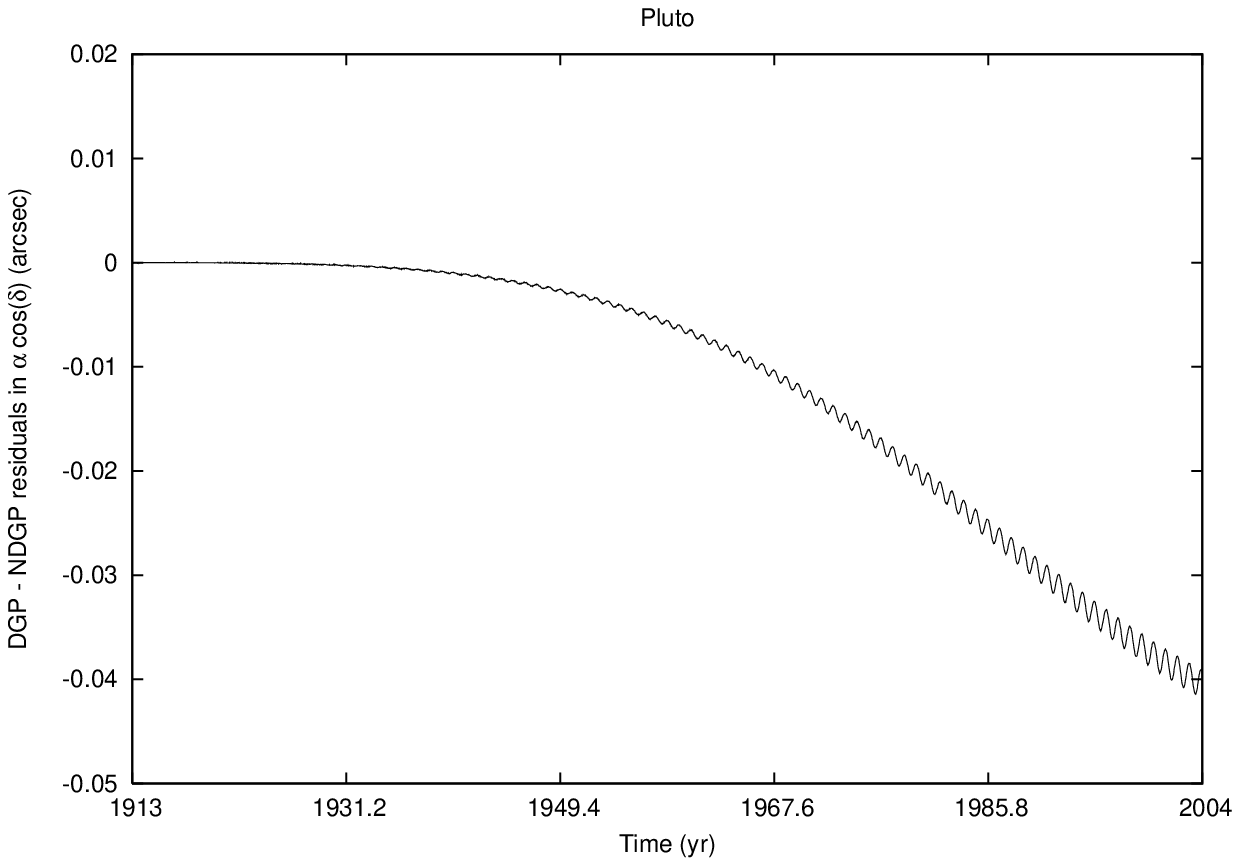}
\end{center}
\caption{\label{pl_ra} Pluto: DGP shift, in arcseconds, for
$\alpha\cos\delta$ over $T=90$ years.}
\end{figure}

\begin{figure}[htbp]
\begin{center}
\includegraphics[width=14cm,height=12cm,angle=0]{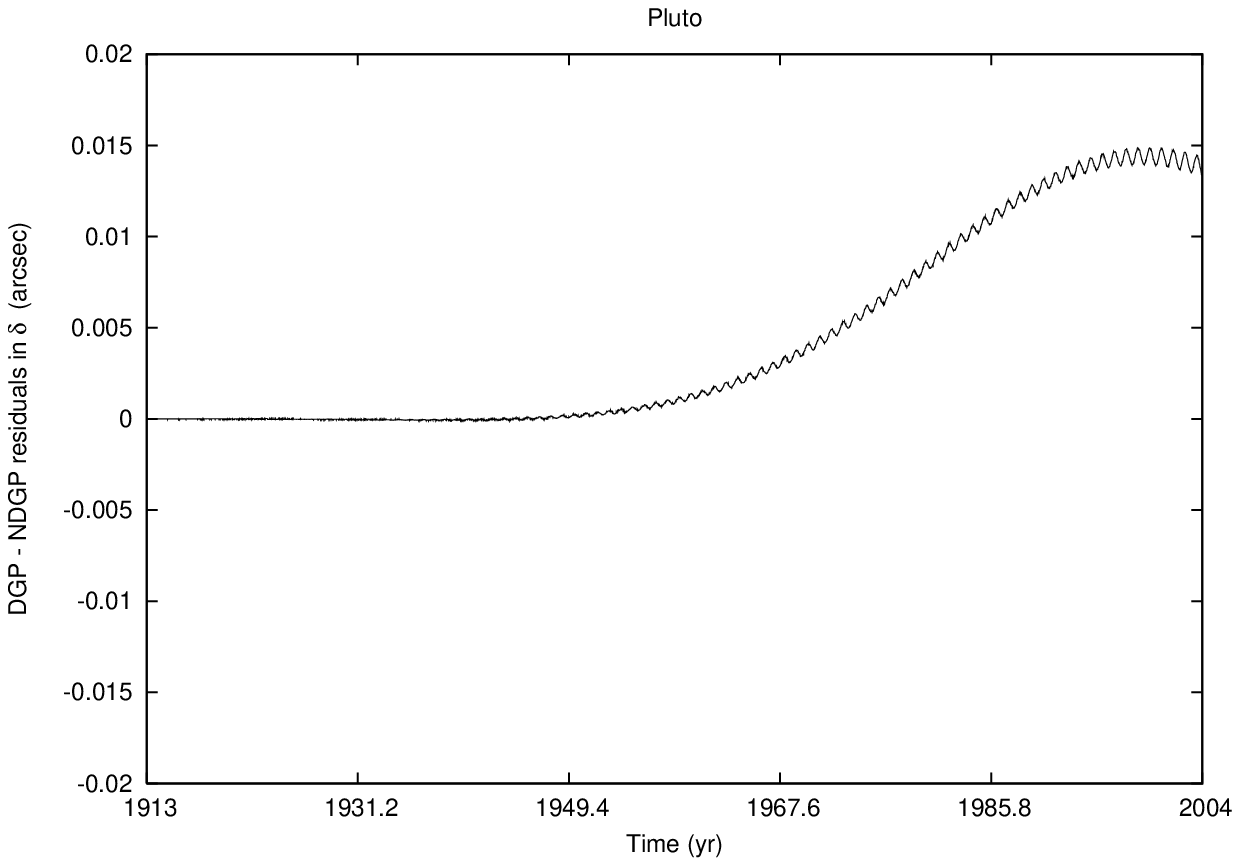}
\end{center}
\caption{\label{pl_de} Pluto: DGP shift, in arcseconds, for
$\delta$ over $T=90$ years.}
\end{figure}

The integration has been performed twice, with and without the DGP
acceleration of \rfr{acc}, and $\alpha\cos\delta$ and $\delta$
have been calculated in both cases. The difference between
$\alpha\cos\delta$ and $\delta$ with and without the perturbation
constitute the final results. They are represented in Figure
\ref{ju_ra}--Figure \ref{pl_de}.

As a control, the same procedure has been performed adopting the
Bulirsh-Stoer (general) method of integration; the results are not
reported here, because the resulting figures are exactly equal to
those here presented (the differences are at level of the sixth
significative digit).

As can be noted, the pattern which would be induced by a DGP
perturbation on the planetary motions cannot be discerned with the
present-day orbital accuracy.

%
%
%


%
%
%
%
%
%
%
%
%
%
\subsection{The role of Cassini and of other future interplanetary missions}
In regard to Cassini, it turns out that recent attempts performed
at JPL to use its first data to improve the orbit of Saturn failed
to reach the required level of accuracy (Jacobson, R.A., private
communication, 2005). Moreover, the duration of the Cassini
mission will only cover one sixth of one entire orbital period of
Saturn. In regard to\footnote{See on the WEB
http://sci.esa.int/science-e/www/area/index.cfm?fareaid=26} GAIA,
it is not tailored for observations of planets: indeed, they are
too bright and cannot be measured exactly. Moreover, as in the
case of Cassini, long stream of observations should be required,
not mere sparse points. Thus, it is doubtful  that any accurate
observations of planets will be obtained from GAIA (Pitjeva, E.V.,
private communication, 2005).

As a consequence, the preferred test-bed for DGP will likely
remain the inner Solar System even in the near future when the
data from the\footnote{See on the WEB
http://sci.esa.int/science-e/www/area/index.cfm?fareaid=30 for
BepiColombo, http://messenger.jhuapl.edu/ for Messenger and
http://sci.esa.int/science-e/www/area/index.cfm?fareaid=64 for
Venus Express. } BepiColombo, Messenger, Venus Express spacecraft
will be available and especially if the currently investigated
projects on the interplanetary laser ranging \cite{Nord03} will be
implemented. Indeed, the use of optical frequencies would
definitely allow to overcome the limitations posed by the solar
corona to the radar ranging technique.

\section{Conclusions}
In this paper we analyzed the possibility of using the outer
planets of the Solar System, especially Jupiter and Saturn, to
test the Dvali-Gabadadze-Porrati multidimensional braneworld model
of gravity via the Lue-Starkman secular precession of perihelion
$\omega$ and the Iorio secular precession of the mean longitude
$\lambda$. Indeed, for the distant planets such effects, which are
independent of the geometrical features of the orbits apart from
second-order terms in the eccentricities $e$, would be  larger
than the other competing perturbations induced by the mismodelled
or unmodelled  Newtonian and Einsteinian features of motion.
Moreover, the current presence of the Cassini spacecraft in the
Saturnian system with its precise radiotechnical apparatus and the
forthcoming astrometric mission GAIA might induce some
expectations about such an intriguing possibility. Unfortunately,
the real situation is less favorable than it was hoped. Indeed,
the investigated new features of motion are currently by one-two
orders of magnitude below the threshold set by the (formal)
accuracy of the most recent determinations of the orbital elements
of Jupiter and Saturn. Moreover, it is doubtful that Cassini will
substantially contribute to improving our knowledge of the orbit
of Saturn to a sufficient level. Indeed, it could yield only
sparse points because its lifetime will span just one sixth of an
entire orbital period of Saturn.

\section*{Acknowledgements}
L.I. thanks R.A. Jacobson (Jet Propulsion Laboratory) and E.V.
Pitjeva (Institute of Applied Astronomy, Russian Academy of
Sciences) for useful correspondence and M. Gasperini (INFN, Bari)
for important references.

\end{document}